# Robust frequency offset estimator for OFDM over fast varying multipath channel

Li Wei, Youyun Xu, Yueming Cai and Xin Xu

This paper presents a robust carrier frequency offset(CFO) estimation algorithm suitable for fast varying multipath channels. The proposed algorithm estimates CFO both in time-domain and frequency-domain using two carefully designed sequences. This novel technique possesses high accuracy as well as large estimation range and works well in fast varying channels.

*Introduction:* Frequency offset estimation is one of the most important task in OFDM receiver. Various frequency estimators have already been developed in the literature, including blind 、semi-blind and data-aided algorithms. Showing great superiority in estimation accuracy、speed and complexity, data-aided algorithms attract much interest in recent years. The normalized frequency offset is usually divided into integral frequency offset(IFO) and fractional frequency offset(FFO), which may be estimated separately or jointly. Most data-aided algorithms have a limited estimation range and can only be used to recover FFO. Morelli proposed a time-domain preamble based algorithm, which utilized an OFDM symbol with multiple identical parts and can estimate the integral and fractional parts simultaneously [1]. However, the algorithm's estimation range is restricted when a satisfactory accuracy is required. In [2]~[4], the differential information between two consecutive OFDM symbols is used to estimate IFO by assuming that the channel is invariable over at least two OFDM symbols duration. This condition is not met any more over fast varying multipath channel.

In this Letter, two special CAZAC sequences are firstly designed for preamble purpose. The fractional part of CFO is then estimated and compensated in time-domain exploiting auto-correlation property of received preamble. Correlating two compensated preamble with local referential preamble in frequency-domain and making full use of the information of the locations corresponding to two correlation peaks, we may obtain the estimation of IFO. Due to the independence in searching for the peak location, the channel is permitted to vary independently from symbol to symbol. This trait makes the new method more suitable for fast varying channel compared with the method in [2]~[4].

*Signal model:* In this letter, we adopt two CAZAC sequences as preamble for it's perfect correlation property [5]. The first sequence is used for FFO estimation. The first and the second sequence are both utilized for IFO estimation after compensation. The CAZAC sequence in time-domain is in the form:

$$x(n) = \exp\left(\frac{j\pi r n^2}{N}\right) \qquad n = 0,1\cdots N-1 \tag{1}$$

These two CAZAC sequences have different parameter $r_1$ and $r_2$. After timing recovery and CP removed, the received signal $y(n)$ may be represented as:

$$y(n) = \sum_{m=0}^{L-1} h(m)x(n-m)e^{j(2\pi\varepsilon n/N+\phi_0)} + w(n) \tag{2}$$

where a Rayleigh fading channel model with exponential power delay profile is used.

$$E\left\{|h(l)|^2\right\} = \exp(-l/D) \tag{3}$$

$L$ is the number of path; $w(n)$ is white Gaussian process with zero mean and variance $\sigma_w^2$.

*FFO estimation:* The FFO is estimated by use of the auto-correlation of received signal (noise is not considered):

$$R_t\left(\varepsilon,r\right) = \frac{1}{N-N/r}\sum_{n=0}^{N-1-N/r} y(n+N/r)y^*(n) = \sum_{m=0}^{L-1}|h(m)|^2 e^{j(2\pi\varepsilon/r)} \tag{4}$$

Let $\varepsilon_i$ 、 $\varepsilon_f$ denote the integral and fractional part of $\varepsilon$ ,and $\lfloor \varepsilon_i/r \rfloor$ is the integer part of $\varepsilon_i/r$ , $\varepsilon_r = \left(\varepsilon_i\right)_{\bmod r}$ , $\varepsilon = \lfloor \varepsilon_i/r \rfloor r + \varepsilon_f + \varepsilon_r$ .Then we get the estimation of FFO:

$$\begin{aligned}\hat{\varepsilon}_f &= \frac{r}{2\pi}\arg\left\{R_t\left(\varepsilon,r\right)\right\} = \frac{r}{2\pi}\left[\arg\left\{R_t\left(\varepsilon_f,r\right)\right\} + \arg\left\{R_t\left(\varepsilon_r,r\right)\right\}\right] \\ &= \begin{cases}\varepsilon_f + \varepsilon_r & 0 \le |\varepsilon_r| < r/2 \\ \varepsilon_f + \varepsilon_r - r & r/2 \le \varepsilon_r < r \\ \varepsilon_f + \varepsilon_r + r & -r < \varepsilon_r \le -r/2\end{cases}\end{aligned} \tag{5}$$

We note that the result is ambiguous by multiples of subcarrier spacing. This ambiguity will be eliminated by IFO estimator and has no influence on the whole estimation.

*IFO estimation:* It is known that the DFT of a CAZAC sequence is also a CAZAC sequence. The IFO estimator is on the basis of this fact. Defining $\mathrm{X(k)}=\left\{X(0)\cdots X(N\text{-}1)\right\}=FFT\left(\mathrm{x(n)}\right)$ 、 $\mathrm{x(n)}=\left\{x(0)\cdots x(N\text{-}1)\right\}$ , we have:

$$y'(n)=y(n)e^{-j2\pi\hat{\varepsilon}_f n/N} \qquad 0\le n\le N-1 \tag{6}$$

$$Z(k)=\frac{1}{\sqrt{N}}\sum_{n=0}^{N-1}y'(n)e^{-j2\pi k\frac{n}{N}} \tag{7}$$

$$\begin{aligned} R_f\left(\varepsilon-\hat{\varepsilon}_f,r,\tau\right)&=\frac{1}{N}\sum_{k=0}^{N-1}X\left(k-\tau\right)_{\mathrm{mod}\,N}Z^*(k)\\ &=\begin{cases}\displaystyle\sum_{m=0}^{L-1}h^*(m)\delta\left(\tau-\left(\left(\varepsilon-\hat{\varepsilon}_f\right)-rm\right)_{\mathrm{mod}\,N}\right)e^{-j\frac{\pi r}{N}m^2-j\phi_0} & \varepsilon-\hat{\varepsilon}_f>0\\ \displaystyle\sum_{m=0}^{L-1}h^*(m)\delta\left(\tau-\left(\varepsilon-\hat{\varepsilon}_f\right)+rm-N\right)e^{-j\frac{\pi r}{N}m^2-j\phi_0} & \varepsilon-\hat{\varepsilon}_f<0\end{cases} \end{aligned} \tag{8}$$

When the parameter $r_1$ and $r_2$ of the CAZAC sequence satisfy the relation $r_1L\le r_2<N/L$, all the local maximal locations of $R_f\left(\varepsilon-\hat{\varepsilon}_f,r_1,\tau\right)$ and $R_f\left(\varepsilon-\hat{\varepsilon}_f,r_2,\tau\right)$ can be distinguished from each other(shown in Fig.1). This is the main idea of IFO estimator. Considering all possible relationship between the locations of two global maximums, we may depict the algorithm as follows:

1. Find the peak location of the two correlations respectively.

   $L_{1\max}=\arg\max\limits_{\tau}\left\{\left|R_f\left(\varepsilon-\hat{\varepsilon}_f,r_1,\tau\right)\right|\right\}$ , $L_{2\max}=\arg\max\limits_{\tau}\left\{\left|R_f\left(\varepsilon-\hat{\varepsilon}_f,r_2,\tau\right)\right|\right\}$
2. If $L_{1\max}<N/2$ , go to step 3, else go to step 4.
3. While$\left(L_{2\max}\ge N/2\right)$ $\left\{L_{2\max}=\left(L_{2\max}+r_2\right)_{\mathrm{mod}\,N}\right\}$; While$\left(L_{2\max}<L_{1\max}\right)$ $\left\{L_{2\max}=\left(L_{2\max}+r_2\right)\right\}$. Go to step 5.
4. While$\left(L_{2\max}<L_{1\max}\right)$ and$\left(L_{2\max}>r_2\right)$ $\left\{L_{2\max}=\left(L_{2\max}+r_2\right)\right\}$.
5. $\tilde{\varepsilon}_i=\left(L_{2\max}\right)_{\mathrm{mod}\,N}$; If $\tilde{\varepsilon}_i>N/2$ , $\tilde{\varepsilon}_i=\tilde{\varepsilon}_i-N$ (assuming $\left|\varepsilon\right|<N/2$ ).

The estimation range of this estimator reaches the whole signal bandwidth .

*Simulation and discussion:* Simulations have been run under the following conditions: $r_1=2$ , $r_2=8$ , $\varepsilon=20$ , Rayleigh fading channel with $L=4$ and $E\left\{\left|h(l)\right|^2\right\}=\exp(-l/2)$ . Channel is assumed to be independent for different OFDM symbol. Both $N=64$ and $N=128$ are simulated. Fig.1 shows that the peaks of two correlations can be identified by location, which forms the foundation of our IFO estimator. In order to illustrate the performance of IFO estimator, we simulate the failure probability of IFO estimator in the absence of FFO estimator and compare it with the failure probability using SCA method in [2]. The failure probability $P_f$ is defined as: $P\{\tilde{\varepsilon}_i \neq \varepsilon_i\}$. We refer to the channel that varies from OFDM symbol to symbol but remains unchanged during one OFDM symbol as varying channel. From Fig. 2, it is obvious that the proposed method is superior to SCA algorithm in terms of failure probability for both varying channel and static channel.

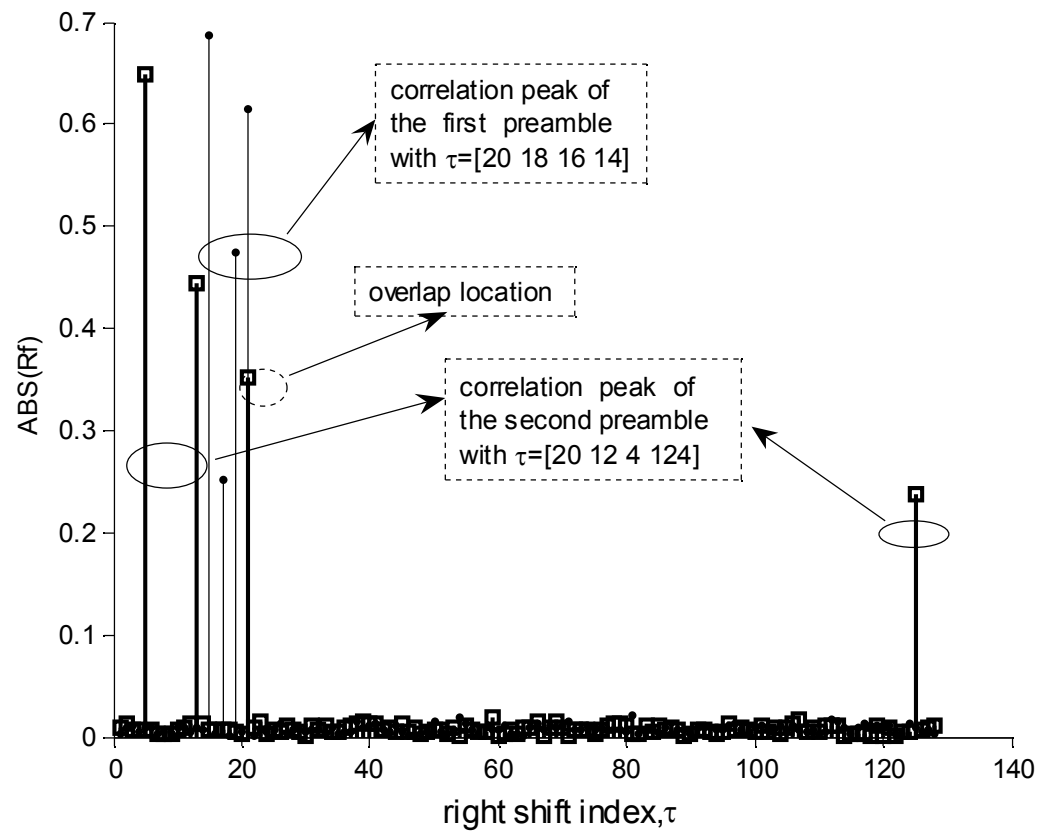

Fig.1 *Relation between two correlations* $\left|R_f\left(\varepsilon-\hat{\varepsilon}_f, r_1, \tau\right)\right|$ 、 $\left|R_f\left(\varepsilon-\hat{\varepsilon}_f, r_2, \tau\right)\right|$ *in peak location with N=128*、*SNR=20dB* *and* $\varepsilon-\hat{\varepsilon}_f=20$

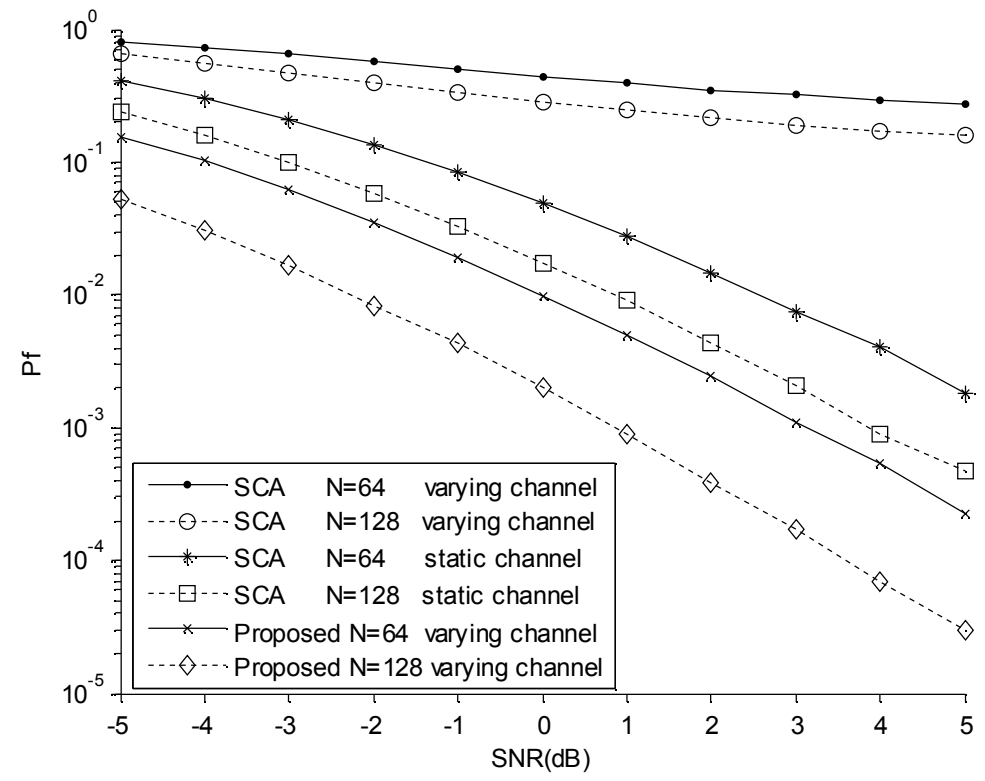

Fig.2 *Failure probability comparison between our method and SCA algorithm.*

Li Wei, Youyun Xu, Yueming Cai and Xin Xu ( Department of mobile communication, Institute of Communications Engineering, PLA University of Science and Technology, Nanjing, China )
E-mail: wlnb@hotmail.com

**References**

1. Morelli, M., and Mengali, U.: 'An improved frequency offset estimator for OFDM applications', *IEEE Commun. Lett*, 1999, 3, pp.75–77
2. Schmidl, T.M., and Cox, D.C.: 'Robust frequency and timing synchronization for OFDM', *IEEE Trans. Commun.*, 1997, 45,pp. 1613–1621
3. Morelli, M., D'Andrea, A. N., and Mengali, U.: 'Frequency ambiguity resolution in OFDM systems', *IEEE Commun. Lett.*, 2000,4, pp. 134–136
4. Chen, C., and Li, J.: 'Maximum likelihood method for integer frequency offset estimation of

OFDM systems’, *Electronics Letters* , 2004,40, (13), pp. 813-814

5. Bomer, L., and Antweiler, M.: ‘Perfect N-phase sequences and arrays’, *IEEE J. Select. Areas Commun*., 1992,10, (4), pp.782-789